\documentclass[english,smallextended]{svjour3}       % onecolumn (second format)
\smartqed  % flush right qed marks, e.g. at end of proof

\usepackage{graphicx}
\usepackage{amssymb}
\usepackage{amsmath}
\usepackage{color}
\begin{document}

\title{Semiconducting double-dot exchange-only qubit dynamics in presence of magnetic and charge noises}

\titlerunning{Semiconducting DEOQ dynamics in presence of magnetic and charge noises}        % if too long for running head

\author{E. Ferraro\and M. Fanciulli\and M. De Michielis}


\institute{E. Ferraro 
              \and 
              M. De Michielis              
             \at CNR-IMM, Unit of Agrate Brianza, Via C. Olivetti 2, 20864 Agrate Brianza (MB), Italy \\
             \email{elena.ferraro@mdm.imm.cnr.it}           
               \and  
              M. Fanciulli \at 
              CNR-IMM, Unit of Agrate Brianza, Via C. Olivetti 2, 20864 Agrate Brianza (MB), Italy\at 
              Dipartimento di Scienza dei Materiali, University of Milano Bicocca, Via R. Cozzi 55, 20125 Milano, Italy\\
              }

\date{Received: date / Accepted: date}

\maketitle

\begin{abstract}
The effects of magnetic and charge noises on the dynamical evolution of the double-dot exchange-only qubit (DEOQ) is theoretically investigated. The DEOQ consisting of three electrons arranged in an electrostatically defined double quantum dot deserves special interest in quantum computation applications. Its advantages are in terms of fabrication, control and manipulation in view of implementation of fast single and two qubit operations through only electrical tuning. The presence of the environmental noise due to nuclear spins and charge traps, in addition to fluctuations in the applied magnetic field and charge fluctuations on the electrostatic gates adopted to confine the electrons, is taken into account including random magnetic field and random coupling terms in the Hamiltonian. The behavior of the return probability as a function of time for initial conditions of interest is presented. Moreover, through an envelope-fitting procedure on the return probabilities, coherence times are extracted when model parameters take values achievable experimentally in semiconducting devices.

\keywords{Qubit architectures\and Quantum dots \and Noise\and Coherence time}
 \PACS{03.67.Lx, 73.21.La, 03.65.Yz}
\end{abstract}

\section{Introduction}
Confinement of electron spins in solid state architectures represents a fruitful platform for universal quantum computation as witnessed by several experimental \cite{Shulman-2012,Veldhorst-2014,Pla-2012,Maune-2012,Bluhm-2011,Tyryshkin-2012} and theoretical \cite{RuiLi-2012,Coish-2005,Shen-2000} proposals. The approaches developed range from quantum dot (QD) \cite{Morton-2011,Veldhorst-2014,Kawakami-2014} to donor-atom nuclear or electron spins \cite{Klymenko-2015,Gamble-2015,Saraiva-2015}. The reasons that make semiconductor nanostructures based qubits an attractive scenario for technological applications are due to their relatively long coherence times, the easy manipulation and fast gate operations. Thanks to the compatibility with the existing semiconductor electronics industry the DEOQ is directly scalable \cite{Loss-1998,DiVincenzo-2000,Taylor-2005,Laird-2010} . In the framework of QDs qubits several architectures have been proposed based on single \cite{Loss-1998}, double \cite{Taylor-2005,Levy-2002,Petta-2005} and triple \cite{DiVincenzo-2000} QDs, implemented in III-V compounds such as GaAs \cite{Kikkawa-1998,Amasha-2008,Koppens-2008,Barthel-2010}, and group IV element like Si \cite{Tyryshkin-2003,Morello-2010,Simmons-2011,Pla-2012,Xiao-2010} but also in InSb \cite{vandenBerg-2013} nanostructures. With the aim to devise an architecture capable to assure the best compromise among fabrication, tunability, fast gate operations, manipulability and scalability, double-dot exchange-only qubit (DEOQ) has been proposed \cite{Shi-2012,Koh-2012}, demonstrated \cite{Kim-2014} and constantly developed \cite{Kim-2015,Thorgrimsson-2017,Prati-npj}. Exchange interactions between adjacent spins suffice for all one- and two-qubit operations \cite{DiVincenzo-2000}. 

The fidelity in the realization of quantum gates is deeply influenced by the unavoidable environmental noise mainly due to two different sources of disturbance that cause decoherence. One contribution to decoherence comes from the magnetic field noise due to the nuclear spins in the host material, in addition to fluctuations in the applied magnetic field needed to remove spin degeneracy of quantum states. We point out that DEOQ is protected against global magnetic fluctuations since it is a decoherence-free subspace qubit. This means that it is only affected by local fluctuations, such as the Overhauser field. The second major disturbance is represented by the charge noise that originates from charge fluctuations on nearby impurities that act as traps or on the electrostatic gates adopted to confine the electrons, that affects the exchange couplings between the spins. Magnetic noise is very sensitive to the host material under consideration. For this reasons it is of minor entity in Si thanks to the presence of stable isotopes with zero nuclear spins while it becomes considerable for example in GaAs compounds, where its effect could be considerable reduced through dynamical decoupling techniques and nuclear polarization. Nevertheless when Si is taken into account, charge traps and fluctuations in the magnetic field remain two issues to face.  

The aim of the present work is to develop a theoretical study on the effects of both magnetic and charge noises on the dynamics of the DEOQ. The main hypothesis is based on an alternative approach with respect to the usual quantum computing gate operations in which an accurate control over timing and duration of pulses is required in order to perform the desired one or two-qubit gate operations \cite{Ferraro-2017,DeMichielis-2015}. In our study we consider the natural evolution in time of the DEOQ as if the control parameters, that are the external gate voltages, are always turned on and two different sources of noise, namely the charge and magnetic noise, are included. In other words we are studying the "always on" configuration and this choice is taken in order to ease the comparison between our results and the experimental ones due to the simple (constant) inputs to be applied by the experimentalist. Note the clear difference with respect to the study presented in Ref. \cite{DeMichielis-2015} where time-variant gate sequences are calculated to obtain rotations along $\hat{x}$ and $\hat{z}$ axes of the Bloch sphere. 

Having a clear and complete picture of how the different physical mechanisms that give birth to environmental noise on the system affect the dynamics and quantify them represents a fundamental step to progress in the development of a qubit technology.

The paper is organized as follows. Section 2 is devoted to the presentation of the effective Hamiltonian model for the DEOQ and to the derivation of the closed analytical form for the unitary evolution operator that accounts for the dynamical evolution starting from an arbitrary initial condition. Sect. 3 contains the main results about the noise analysis in which two different initial conditions are considered and an estimation on the coherence times is given. Finally in Sect. 4, some concluding remarks are summarized.

\section{Double-dot exchange-only qubit model}
This Section presents the effective Hamiltonian model that accounts for the quantum behaviour of the DEOQ in the regime of low energy excitations. 

\begin{figure}[htbp]
	\begin{center}
		\includegraphics[width=0.3\textwidth]{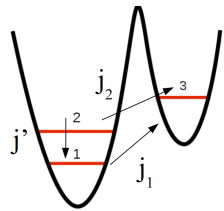}
	\end{center}
	\caption{Schematic energy profile of the double-dot exchange-only qubit. Each level has a twofold spin degeneracy.}\label{hybrid} 
\end{figure}

The DEOQ arises from a new elaboration of the triple-dot exchange-only qubit proposed in Ref. \cite{DiVincenzo-2000} in which the architectures is based on the fabrication of two QDs instead of three. In particular three electrons are distributed during the operations between the two QDs, with at least one electron in each. In Fig. \ref{hybrid} a schematic representation is showed. 

It represents a promising compromise between high speed and simple fabrication for solid state implementations of single qubit and two qubits quantum logic gates. The Schrieffer-Wolff effective Hamiltonian that describes in a simple and compact form the qubit by combining a Hubbard-like model with a projector operator method is derived in Ref. \cite{Ferraro-2014}. As a result, the Hubbard-like Hamiltonian is transformed into an equivalent expression that is the sum of contributions in which each term represents the exchange interaction between each pair of electrons involved, beyond the Zeeman term. The final closed form for the effective Hamiltonian in $\hbar$ units is given by
\begin{equation}
H=\frac{1}{2}E^z(\sigma_{1}^z+\sigma_{2}^z+\sigma_{3}^z)+\frac{1}{4}j'\boldsymbol{\sigma}_{1}\cdot\boldsymbol{\sigma}_{2}+\frac{1}{4}j_1\boldsymbol{\sigma}_{1}\cdot\boldsymbol{\sigma}_{3}+\frac{1}{4}j_2\boldsymbol{\sigma}_{2}\cdot\boldsymbol{\sigma}_{3},
\end{equation} 
where $\boldsymbol{\sigma}_i$ $(i=1,2,3)$ are the Pauli operators referring to each electron and $E^z=g\mu_BB^z$ is the Zeeman energy associated to the magnetic field $\bold{B}$ lying in the $\hat{z}$ direction with $g$ the electron g-factor and $\mu_B$ the Bohr magneton. The exchange couplings $j'$, $j_1$ and $j_2$, whose explicit expressions can be found in Ref. \cite{Ferraro-2014}, include the effects of dot tunneling, dot bias and both on-site and off-site Coulomb interactions. Each coupling term contains the ferromagnetic direct exchange between the two electrons from their Coulomb interactions and the anti-ferromagnetic superexchange. Consequently, in principle the value of each coupling can be either positive or negative and it depends strictly on the values of the parameters. However the superexchange term is usually larger than the direct one, leading to positive values for $j_1$ and $j_2$ \cite{Jefferson-1996}. In general, the coupling constants are tunable thanks to the control on the tunneling couplings which can be provided by external gates and on the inter-dot bias voltage. On the contrary, the Coulomb energy as well as the intra-dot bias voltage, directly linked to the exchange coupling j' \cite{Ferraro-2014}, are geometry dependent and they cannot be easily tuned. These approximations reflect the realistic conditions in which a QD is operated. 

The effective Hamiltonian is projected in the eigenspace spanned by the logical basis introduced in Ref. \cite{Shi-2012}. To encode the DEOQ we restrict to the two-dimensional subspace of three-spin states with spin quantum numbers $S=1/2$ and $S_z=-1/2$. $S$ and $S_z$ represent the total angular momentum state and its projection along $\hat{z}$ respectively. We point out that only states with the same $S$ and $S_z$ can be coupled by spin independent terms in the Hamiltonian. The logical basis $\{|0\rangle,|1\rangle\}$ is constituted by singlet and triplet states of a pair of electrons, for example the pair in the left dot, in combination with the single angular momentum state of the third spin, localized in the right dot, through appropriate Clebsh-Gordan coefficients. This means that the logical states chosen are finally expressed in this way
\begin{equation}\label{01}
|0\rangle\equiv|S\rangle|\!\downarrow\rangle, \qquad |1\rangle\equiv\sqrt{\frac{1}{3}}|T_0\rangle|\!\downarrow\rangle-\sqrt{\frac{2}{3}}|T_-\rangle|\!\uparrow\rangle
\end{equation}
where $|S\rangle$, $|T_0\rangle$ and $|T_-\rangle$ are respectively the singlet and triplet states
\begin{equation}
|S\rangle=\frac{|\!\uparrow\downarrow\rangle-|\!\downarrow\uparrow\rangle}{\sqrt{2}}, \quad |T_0\rangle=\frac{|\!\uparrow\downarrow\rangle+|\!\downarrow\uparrow\rangle}{\sqrt{2}}, \quad |T_-\rangle=|\!\downarrow\downarrow\rangle
\end{equation}
and $|\!\uparrow\rangle$ and $|\!\downarrow\rangle$ denote a single state electron with spin-up and spin-down respectively. Explicit calculations of the matrix elements of the Hamiltonian in this basis give as final result
\begin{equation}\label{effmatrix}
H= \left(\begin{array}{cc}
 -\frac{E^z}{2}-\frac{3}{4}j' &\;\; -\frac{\sqrt{3}}{4}(j_1-j_2) \\
 -\frac{\sqrt{3}}{4}(j_1-j_2) &\;\; -\frac{E^z}{2}+\frac{1}{4}j'-\frac{1}{2}(j_1+j_2)
\end{array}\right).
\end{equation}
The effective Hamiltonian just derived represents the starting point to successfully analyze the dynamical behavior of the system. 

\subsection{Analytical closed form of the evolution operator}
The evolution operator $U(t)=e^{-iHt}$ associated to the Hamiltonian (\ref{effmatrix}) in the $2\times 2$ logical basis $\{|0\rangle,|1\rangle\}$ can be recast in a compact analytical closed form exploiting the Cayley-Hamilton theorem. The matrix exponential is expressible as a polynomial of order n$\--$1 as given by the following identity
\begin{equation}\label{CH}
e^{-iHt}=s_0(t)I_2+s_1(t)(-iH),
\end{equation}
where $I_2$ is the $2\times 2$ identity matrix and 
\begin{equation}
s_0(t)=\frac{\lambda_1 e^{\lambda_2 t}-\lambda_2 e^{\lambda_1 t}}{\lambda_1-\lambda_2},\qquad s_1(t)=\frac{e^{\lambda_1 t}-e^{\lambda_2 t}}{\lambda_1-\lambda_2}.
\end{equation}
The coefficients $\lambda_1$ and $\lambda_2$ are the eigenvalues of the Hamiltonian (\ref{effmatrix}) conveniently multiplied by the imaginary factor $-i$. Eq.(\ref{CH}) after algebraic manipulations has been recast into the following form 
\begin{equation}\label{CH2}
e^{-iHt}=e^{\lambda_1t}I_2+\frac{e^{\lambda_1t}-e^{\lambda_2t}}{\lambda_1-\lambda_2}(-iH-\lambda_1I_2).
\end{equation}
Equation (\ref{CH2}) constitutes a fundamental result exploitable for studying the dynamical behaviour of the DEOQ starting from an arbitrary initial condition and in the always on configuration.

\section{Noise analysis and coherence times}
In this Section the dynamical behaviour of the DEOQ is investigated when different source of noises are included.

Starting from the return probability of finding the qubit in a given logical state when the initial condition is imposed and the system is free to evolve in time, the disorder averaged return probability for two different initial conditions of interest considering both the magnetic disorder and the exchange one is calculated. DEOQ is protected against global magnetic fluctuations and, as it is evident in Eq.(\ref{effmatrix}), the Zeeman term only provides a global energy shift. For this reason the magnetic noise is included in our analysis by local magnetic field fluctuations acting between the two QDs. It is assumed that the disturbance obeys to a Gaussian distribution $f_{\delta E}(\delta E)$ with zero mean and standard deviation $\sqrt{2}\sigma_E$. In an analogous manner the exchange couplings $j_1$ and $j_2$ follow a Gaussian distribution $f_{j_i}(j_i)$, $i=1,2$, restricted to non-negative values with mean $j_{0i}$ and standard deviation $\sigma_{j_i}$. The intra-dot exchange coupling $j'$ as we have pointed out in Section 2 is fixed by the geometry of the system and not tunable from external gates.

Given an arbitrary quantity $P$, that in the framework of our work is represented by the return probability, its disorder average is defined, following an analogous approach developed in \cite{DasSarma-2016,DasSarma-2017,Wu-2017}, by 
\begin{equation}\label{average}
[P]_{\alpha}=\int_0^{+\infty}\int_0^{+\infty}\int_{-\infty}^{+\infty}dj_1dj_2d(\delta E)f_{\delta E}(\delta E)f_{j_1}(j_1)f_{j_2}(j_2)P
\end{equation}
where
\begin{equation}
f_{\delta E}(\delta E)=\frac{1}{2\sigma_E\sqrt{\pi}}e^{-\frac{(\delta E)^2}{4\sigma_E^2}}
\end{equation}
and
\begin{equation}
f_{j_i}(j_i)=\frac{1}{\sigma_{j_i}\sqrt{2\pi}}\frac{2}{1+erf(\frac{j_{0i}}{\sigma_{j_i}\sqrt{2}})}e^{-\frac{(j_i-j_{0i})^2}{2\sigma_{j_i}^2}},
\end{equation}
with $i=1,2$. In the following the results are presented in correspondace to two initial conditions of interest when several values of $\sigma_E$ and $\sigma_{j_1}=\sigma_{j_2}\equiv\sigma_j$ are considered.

\subsection{Initial condition $|\psi(0)\rangle=|0\rangle$}
Let's initialize the qubit in the logical state $|\psi(0)\rangle=|0\rangle$ and follow the dynamical evolution looking at the return probability $P_{|0\rangle}(t)$ when subjected to the noise. 

\begin{figure}[htbp]
	\begin{center}
		\includegraphics[width=\textwidth]{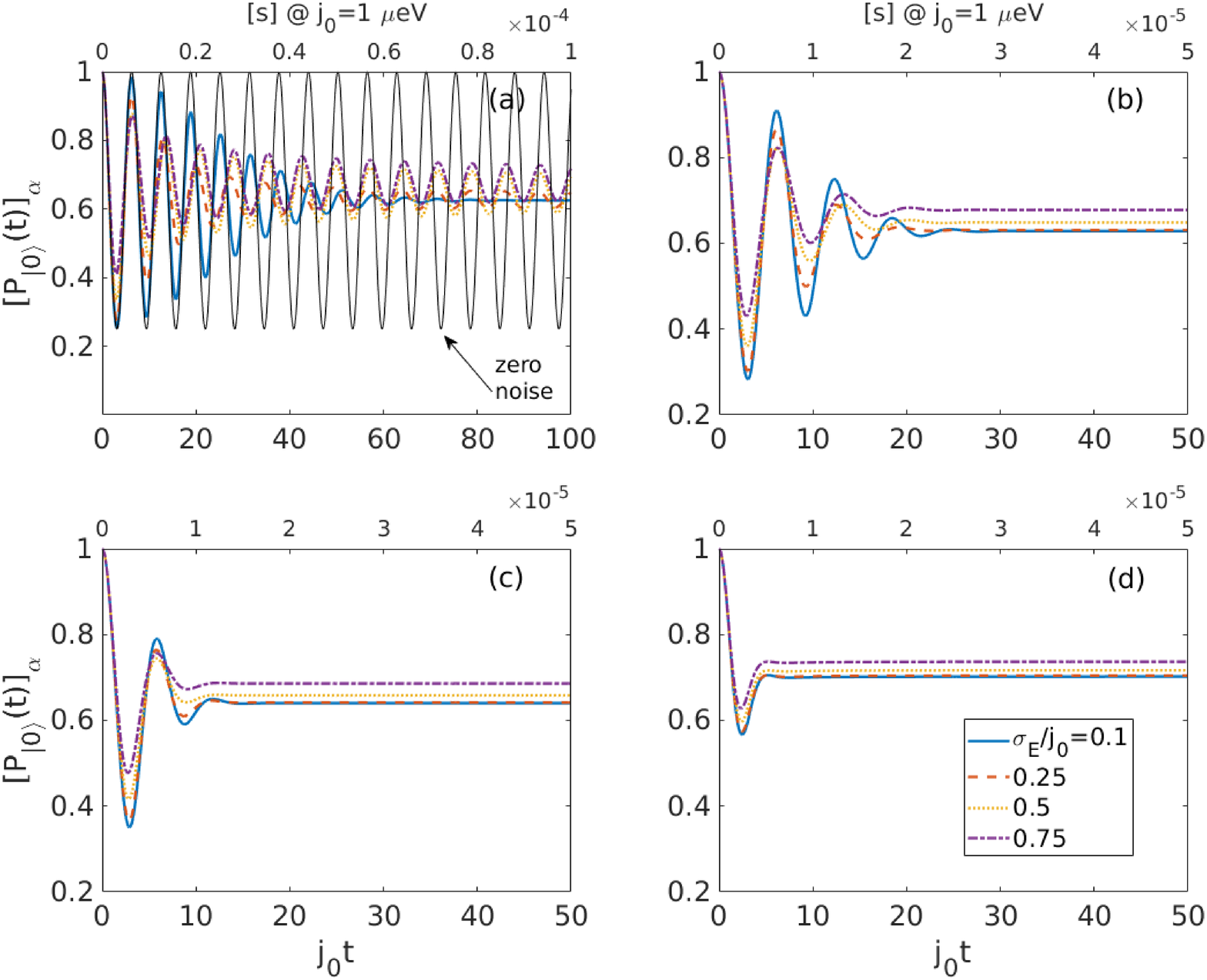}
	\end{center}
	\caption{Plot of the disordered-averaged probability $[P_{|0\rangle}(t)]_{\alpha}$ in correspondence to different values of $\sigma_E$. Each subplot is related to a specific value of $\sigma_j$: (a) $\sigma_j=0$, (b) $\sigma_j=0.1 j_0$, (c) $\sigma_j=0.2 j_0$, (d) $\sigma_j=0.5 j_0$. The second horizontal $\hat{x}$ axis in the upper part of each subplot provides an estimation of the evolution in seconds when $j_0=1 \mu eV$.}\label{cond1_1} 
\end{figure}

After some algebraic manipulations that involve the evolution operator (\ref{CH2}) previously defined applied to the initial condition fixed, the probability $P_{|0\rangle}(t)$ is finally given by
\begin{equation}\label{P}
P_{|0\rangle}(t)=1-\frac{4C^2}{(A-B)^2+4C^2}\sin^2(\beta t),
\end{equation}
where 
\begin{align}\label{coeff1}
&A=\frac{E^z}{2}+\frac{3}{4}j',\nonumber\\
&B=\frac{E^z}{2}-\frac{1}{4}j'+\frac{1}{2}(j_1+j_2),\nonumber\\
&C=\frac{\sqrt{3}}{4}(j_1-j_2)
\end{align}
and
\begin{equation}\label{coeff2}
\beta=\frac{\sqrt{(A-B)^2+4C^2}}{2}.
\end{equation}
The multiple integral for $[P_{|0\rangle}(t)]_{\alpha}$ obtained inserting Eq.(\ref{P}) into Eq.(\ref{average}) must be determined numerically. The parameters chosen for the following numerical calculation are given in function of $j_0$, that can be extracted doing specific calculations through a simulator based on spin density functional theory and mainly depends on the geometrical parameters and on the material of the physical qubit under investigation \cite{DeMichielis-2015}. With this choice our results are completely general and easily exportable to whatever physical context once that $j_0$ is calculated. The parameters fixed in all the following figures are the mean values for the two Gaussian distributions followed by the exchange couplings $j_1$ and $j_2$ that are respectively $j_{01}=0.5 j_0$, $j_{02}=1.5 j_0$, the fixed intra-dot exchange coupling $j'=0.5 j_0$ and the external magnetic field $E^z=10 j_0$. Fig. \ref{cond1_1} reports the results when $\sigma_E$ is varied in correspondence to different values of $\sigma_j$. It is shown that the return probability oscillates around and decays to a steady-state value. The decay rate as expected is set by the disorder strength quantified by the parameters $\sigma_E$ and $\sigma_j$ and increases when they increase. The steady-state value also experiences sensible modifications when magnetic disorder is included. Moreover, in panel (a), the dynamical behavior with zero noise, i.e. $\sigma_E=\sigma_j=0$ (black line), is reported. Those oscillations are not the common Larmor nor Rabi ones, due to the fact that the rotation is not performed along the $\hat{z}$ axis nor an axis in the $\hat{x}$-$\hat{y}$ plane of Bloch sphere.

\begin{figure}[htbp]
	\begin{center}
		\includegraphics[width=\textwidth]{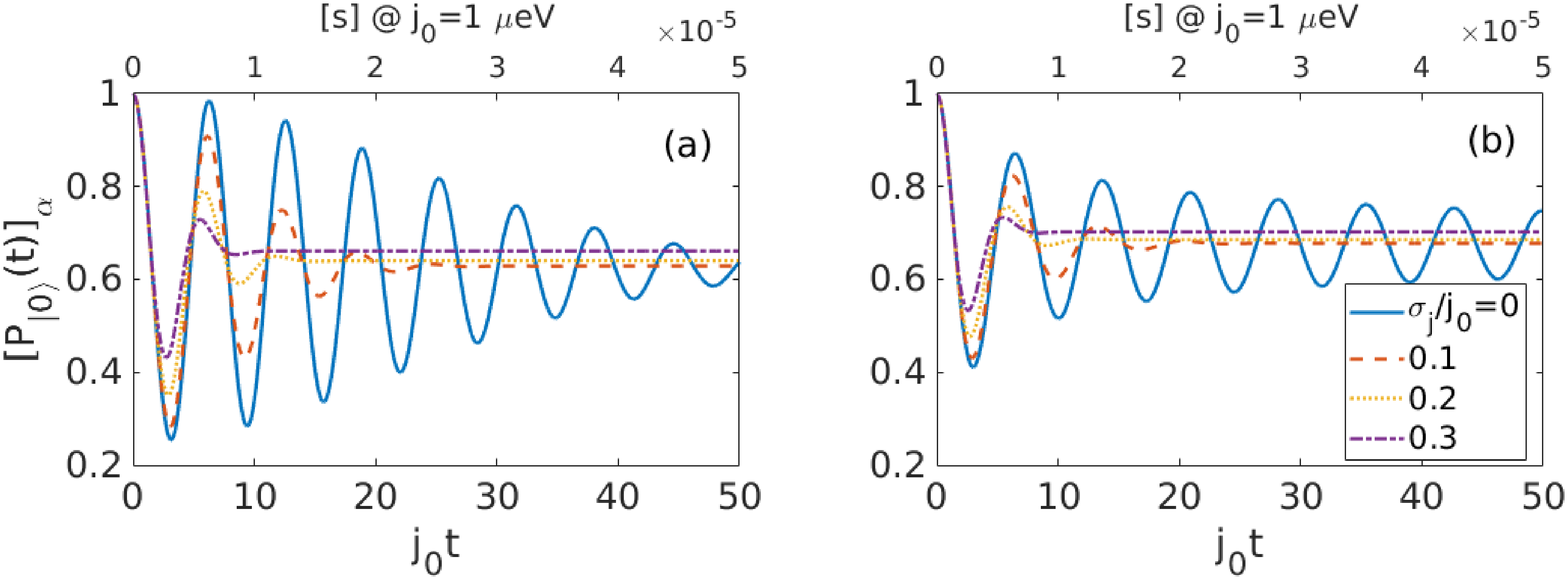}
	\end{center}
	\caption{Plot of the disordered-averaged probability $[P_{|0\rangle}(t)]_{\alpha}$ in correspondence to different values of $\sigma_j$. Each subplot is related to a specific value of $\sigma_E$: (a) $\sigma_E=0.1 j_0$, (b) $\sigma_E=0.75 j_0$.The second horizontal $\hat{x}$ axis in the upper part of each subplot provides an estimation of the evolution in seconds when $j_0=1 \mu eV$.}\label{cond1_2} 
\end{figure}

Fig. \ref{cond1_2} reports the results when $\sigma_j$ is varied in correspondence to two fixed values of $\sigma_E$. In both cases, charge noise is very effective in dampening the oscillations. 

\begin{figure}[htbp]
	\begin{center}
		\includegraphics[width=\textwidth]{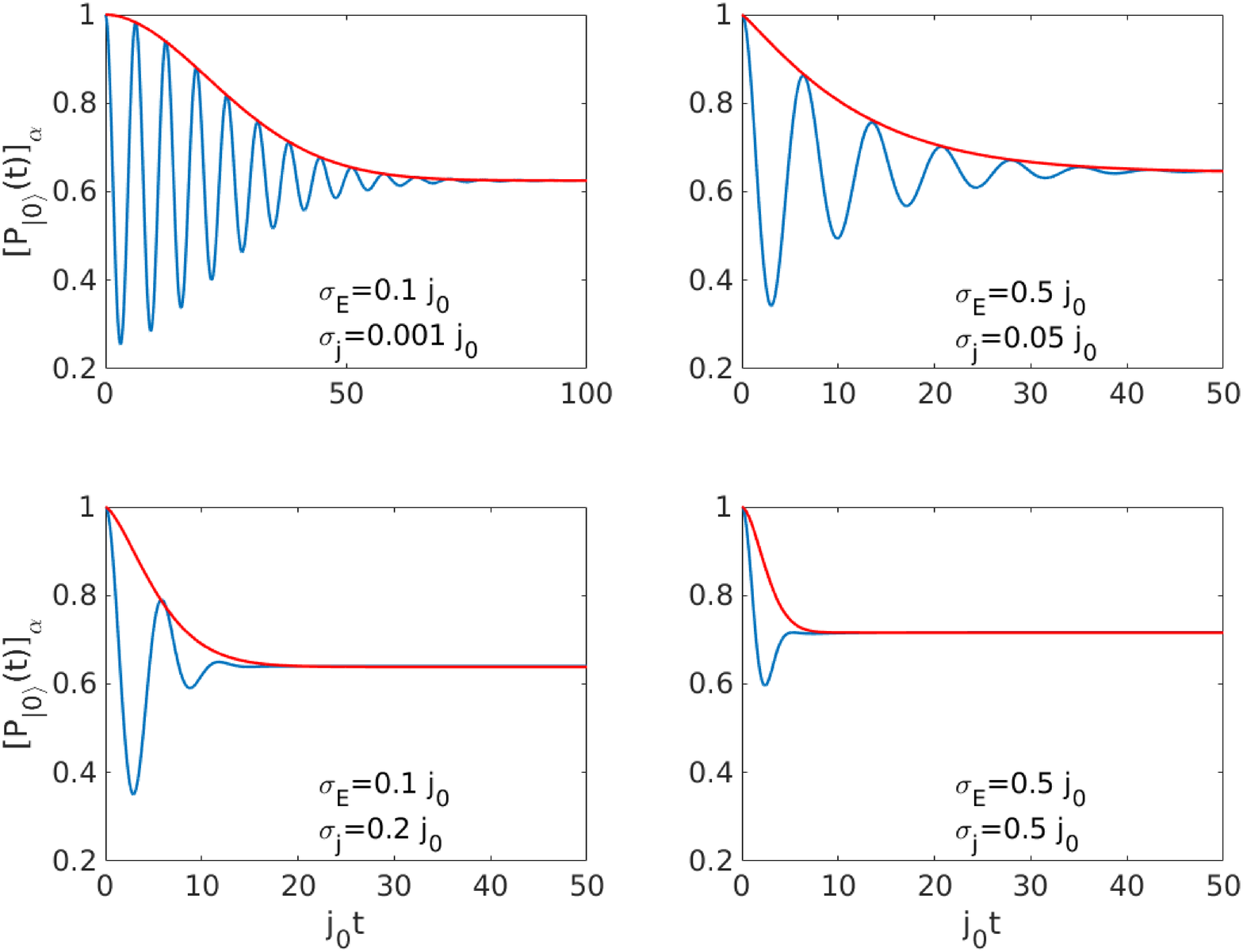}
	\end{center}
	\caption{Examples for the disorder-averaged return probability $[P_{|0\rangle}(t)]_{\alpha}$ (blue) for various system parameters. The red lines show the least-squares fit of the upper envelope of $[P_{|0\rangle}(t)]_{\alpha}$ to the exponential decay form in Eq. (\ref{fit_eq}).}\label{fit} 
\end{figure}

\begin{figure}[htbp]
\centering
\includegraphics[width=5.5cm]{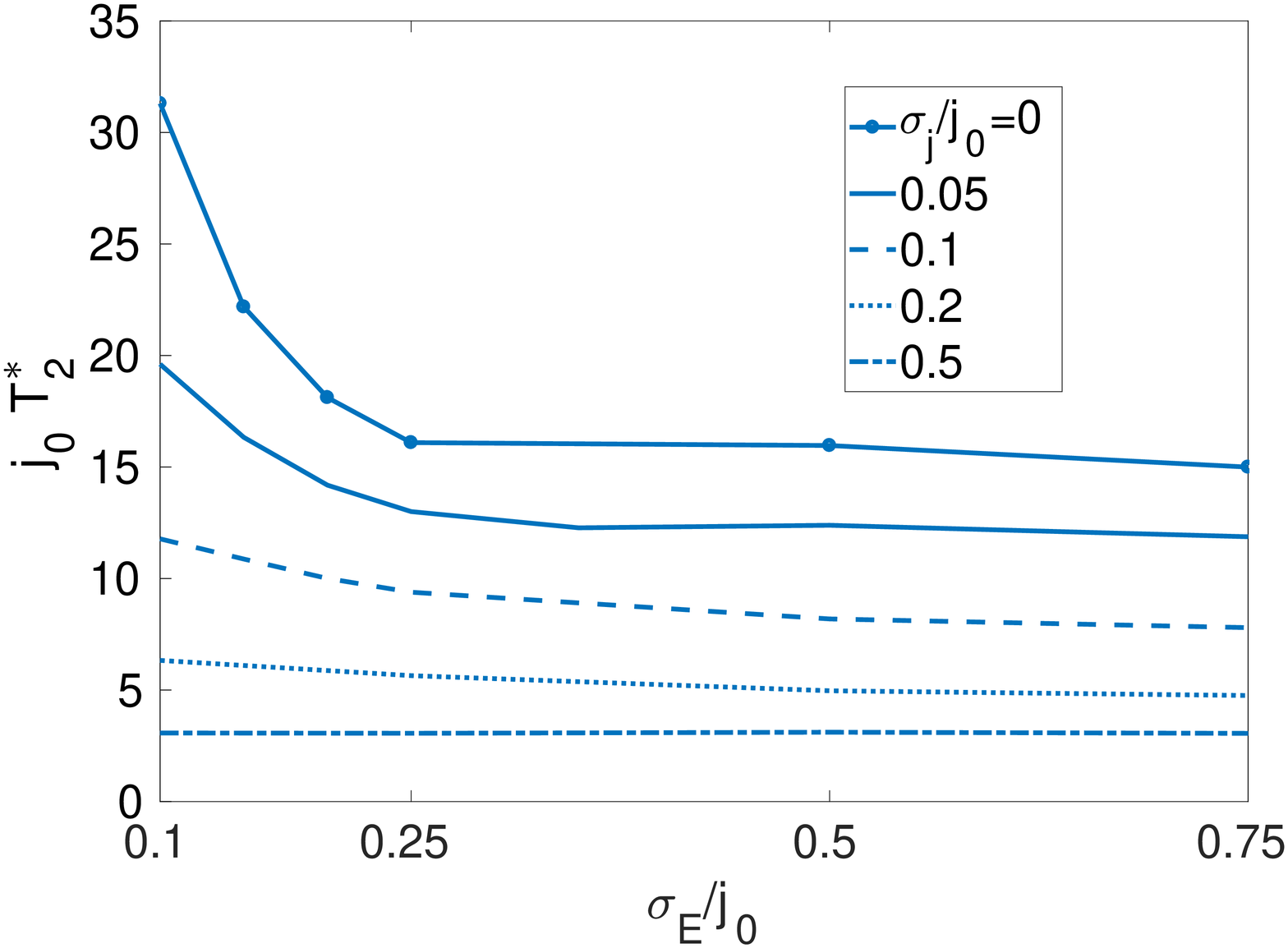}\quad\includegraphics[width=5.5cm]{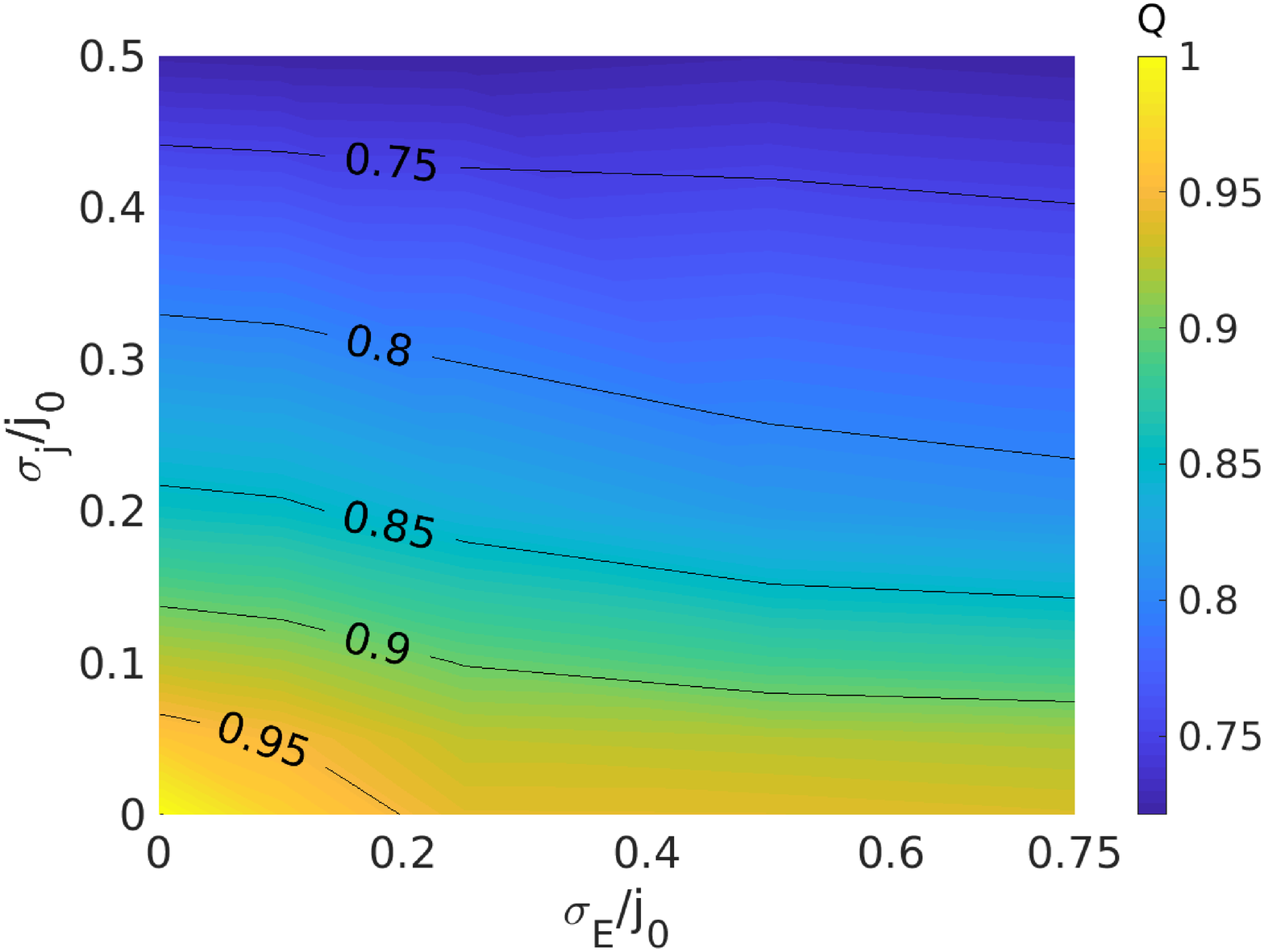}
\caption{Left: Plot of the dimensionless coherence time $j_0 T_2^{\ast}$ as a function of $\sigma_E/j_0$ for different values of $\sigma_j/j_0$. Right: Quality factor as a function of $\sigma_E$ and $\sigma_j$, the black contour lines mark some significant levels of Q.}\label{Q_cond1} 
\end{figure}

In the following starting from the return probabilities just derived the intrinsic coherence time $T_2^{\ast}$ is estimated. From an experimental point of view, a common way to define $T_2^{\ast}$ is trough a Ramsey type experiment \cite{Thorgrimsson-2017}. Such experiment requires a precise sequence starting with a $\pi/2$ pulse that bring the $|0\rangle$ state to the superposition $\frac{1}{\sqrt{2}}(|0\rangle+|1\rangle)$, then the qubit is free to evolve for a time T and finally a second $\pi/2$ pulse is applied. In this way the coherence time $T_2^{\ast}$ is extracted by fitting the decay of the oscillation. Our approach is instead based on the "always on" configuration, that means that the external parameters of control, i.e. gate voltages, are always turned on \cite{DasSarma-2017,Wu-2017} during the qubit evolution. This definition is based on constant amplitudes of the control signals so it is independent on their detailed time behavior. In particular the figure of merit is represented by the dimensionless quantity $j_0 T_2^{\ast}$. It appreciates the number of coherent oscillations shown by the return probability before it decays. In order to proceed with this analysis, the numerical results just presented are exploited. The main idea is to extract the values of the coherence times from the return probabilities through an envelope-fitting procedure. Firstly the envelope of the return probability is derived, then we look for a curve of the form
\begin{equation}\label{fit_eq}
P(t)=P(0)+\left(1-P(0)\right)e^{-(t/T_2^{\ast})^{\alpha}}
\end{equation}
that closely approximates the envelope function previously derived. 
Fig. \ref{fit} shows some explicative examples of the decaying $[P_{|0\rangle}(t)]_{\alpha}$ curves in correspondence to different noise parameters and highlights the good fit of Eq.(\ref{fit_eq}) (red lines) to the upper envelope.
The fitting parameters obtained give an estimation on the dimensionless coherence times as reported in the left part of Fig. \ref{Q_cond1} for different values of $\sigma_j$. 
The clear trend of $j_0T_2^{\ast}$ shows that it decreases increasing either type of disorder.
The magnetic noise affects lesser the oscillatory behaviour of the return probability, that on the contrary is deeply influenced by the entity of the charge noise. 

A function that gives an immediate measure of the coherence of the qubit in an intuitive and clear way is the quality factor
\begin{equation}
Q=\exp\left(-\frac{1}{j_0 T_2^{\ast}}\right),
\end{equation}
that represents the exponential decay factor for the return probability over $1/j_0$. In the right part of Fig. \ref{Q_cond1} a two-dimensional map in which the dependence of $Q$ on the charge noise $\sigma_j$/$j_0$ and on the magnetic noise $\sigma_E$/$j_0$ is reported. When both the sources of noise are suppressed, i.e. $\sigma_j=\sigma_E=0$, the maximum of $Q$ corresponding to the value of 1 is found. As it can be seen, $Q$ factor is strongly reduced by charge noise whereas is less sensitive to magnetic noise. 

\begin{figure}[htbp]
	\begin{center}
		\includegraphics[width=0.55\textwidth]{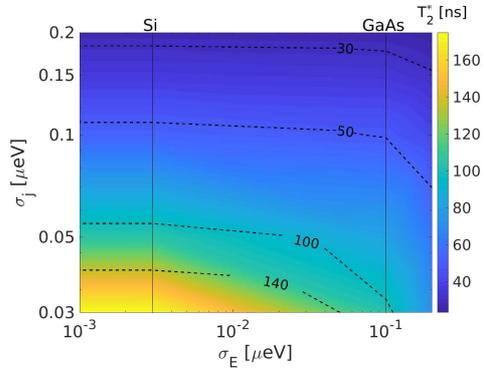}
	\end{center}
	\caption{2D log-log plot of $T_2^{\ast}$ as a function of $\sigma_E$ and $\sigma_j$ when $j_0=1 \mu eV$. The black dashed contour lines mark some levels of $T_{2}^{\ast}$ and vertical solid ones typical minimum magnetic noise variances for Si and GaAs.}\label{3D_cond1} 
\end{figure}

The results obtained in normalized units are very general and mainly depends on $j_0$. This parameter strongly depends on the geometry and on the nature of the material composing the qubit and in order to convert the normalized units in physical ones it is necessary to explicate it. We consider a DEOQ achieving a reasonable value of $j_0$=1 $\mu eV$ \cite{DeMichielis-2015} and including this value in our analysis it is possible to extract a range of physical coherence times for the DEOQ in presence of environmental noise that goes from units to tens of $j_0T_2^{\ast}$ that corresponds to tens up to hundreds of ns. In Fig. \ref{3D_cond1} a 2D map in logarithmic scale of the resulting coherence times is presented when realistic physical units are considered for Si and GaAs. In this figure, vertical lines highlighting the typical values $\sigma_E$=3 n$eV$ and 100 n$eV$ for Si and GaAs respectively \cite{Mehl-2015-2} are added to mark the minimum $\sigma_E$ due to background nuclear spins in each material. 

\subsection{Initial condition $|\psi(0)\rangle=\frac{1}{\sqrt{2}}(|0\rangle+|1\rangle)$}
In an analogous manner this subsection reports the results obtained when the qubit is prepared in an initial condition that is a superposition of the two logical states $|\psi(0)\rangle=\frac{1}{\sqrt{2}}(|0\rangle+|1\rangle)$, for which the analytical expression for the return probability of finding the DEOQ in the logical state $|0\rangle$ is given by
\begin{equation}
P_{|0\rangle}(t)=\frac{1}{2}\left[1+\frac{4C(A-B)}{(A-B)^2+4C^2}\sin^2(\beta t)\right],
\end{equation}
where the parameters involved $A,B,C$ and $\beta$ are defined in Eqs.(\ref{coeff1}) and (\ref{coeff2}).

\begin{figure}[htbp]
	\begin{center}
		\includegraphics[width=\textwidth]{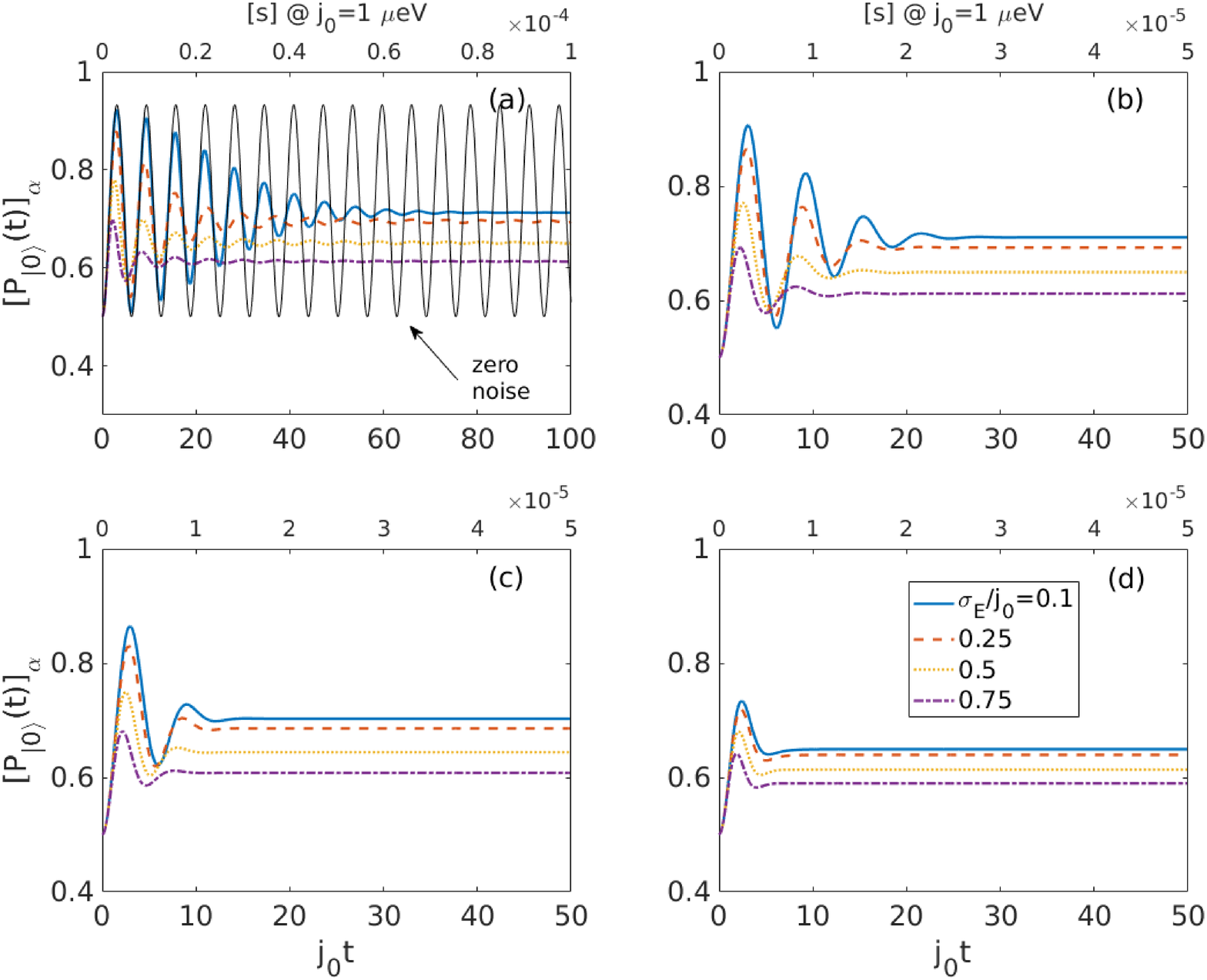}
	\end{center}
	\caption{Plot of the disordered-averaged probability $[P_{|0\rangle}(t)]_{\alpha}$ in correspondence to different values of $\sigma_E$. Each subplot is related to a specific value of $\sigma_j$: (a) $\sigma_j=0$, (b) $\sigma_j=0.1 j_0$, (c) $\sigma_j=0.2 j_0$, (d) $\sigma_j=0.5 j_0$. The second horizontal $\hat{x}$ axis in the upper part of each subplot provides an estimation of the evolution in seconds when $j_0=1 \mu eV$.}\label{cond2_1} 
\end{figure}

\begin{figure}[htbp]
	\begin{center}
		\includegraphics[width=\textwidth]{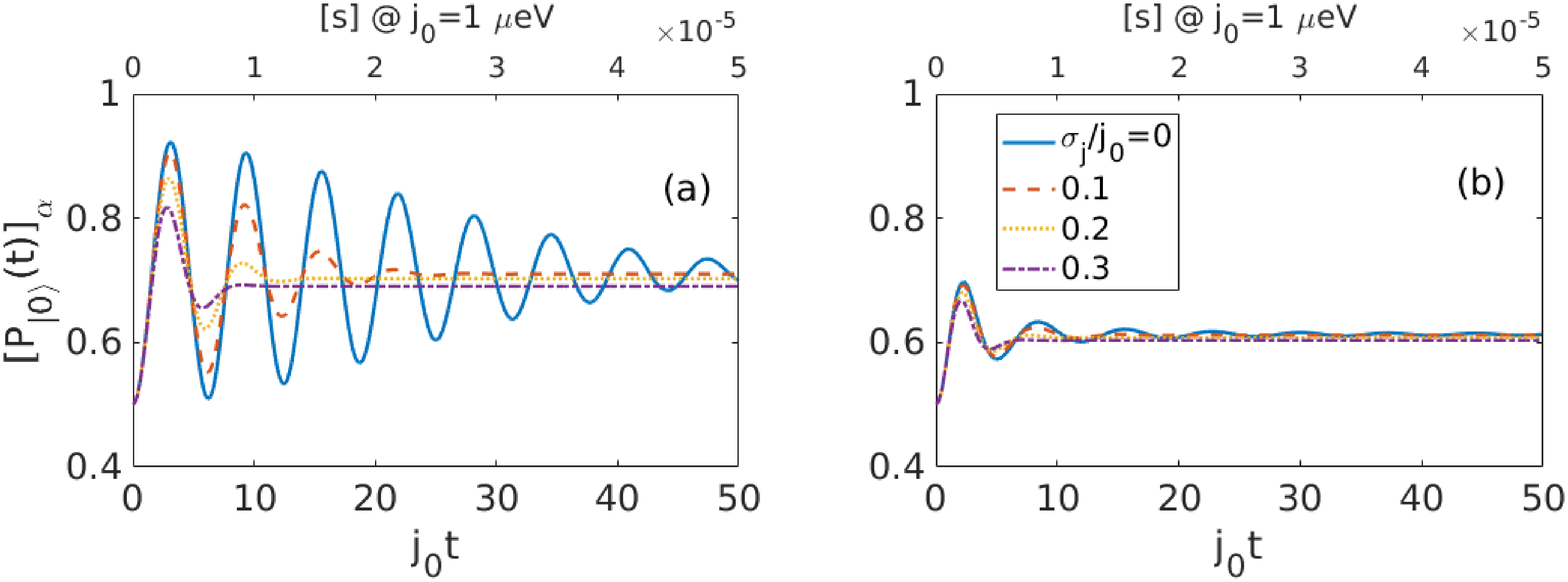}
	\end{center}
	\caption{Plot of the disordered-averaged probability $[P_{|0\rangle}(t)]_{\alpha}$ in correspondence to different values of $\sigma_j$. Each subplot is related to a specific value of $\sigma_E$: (a) $\sigma_E=0.1 j_0$, (b) $\sigma_E=0.75 j_0$. The second horizontal $\hat{x}$ axis in the upper part of each subplot provides an estimation of the evolution in seconds when $j_0=1 \mu eV$.}\label{cond2_2} 
\end{figure}

Once again the expected oscillatory behaviour is reproduced with oscillations gradually reduced when noise increases. We notice that the steady state reached by the qubit, that is larger in correspondence to smaller values of $\sigma_E$ (Fig. \ref{cond2_1}), is more susceptible to the magnetic disorder with respect to the previous case in which the initial condition is the $|0\rangle$ logical state. As done in Fig.\ref{cond1_1}(a), Fig. \ref{cond2_1}(a) displays also the dynamical evolution when zero noise is considered (black line).
Fig. \ref{cond2_2} reports the results on the coherence time when $\sigma_j$ is varied in correspondence to two fixed values of $\sigma_E$. 

\begin{figure}[htbp]
\centering
\includegraphics[width=5.5cm]{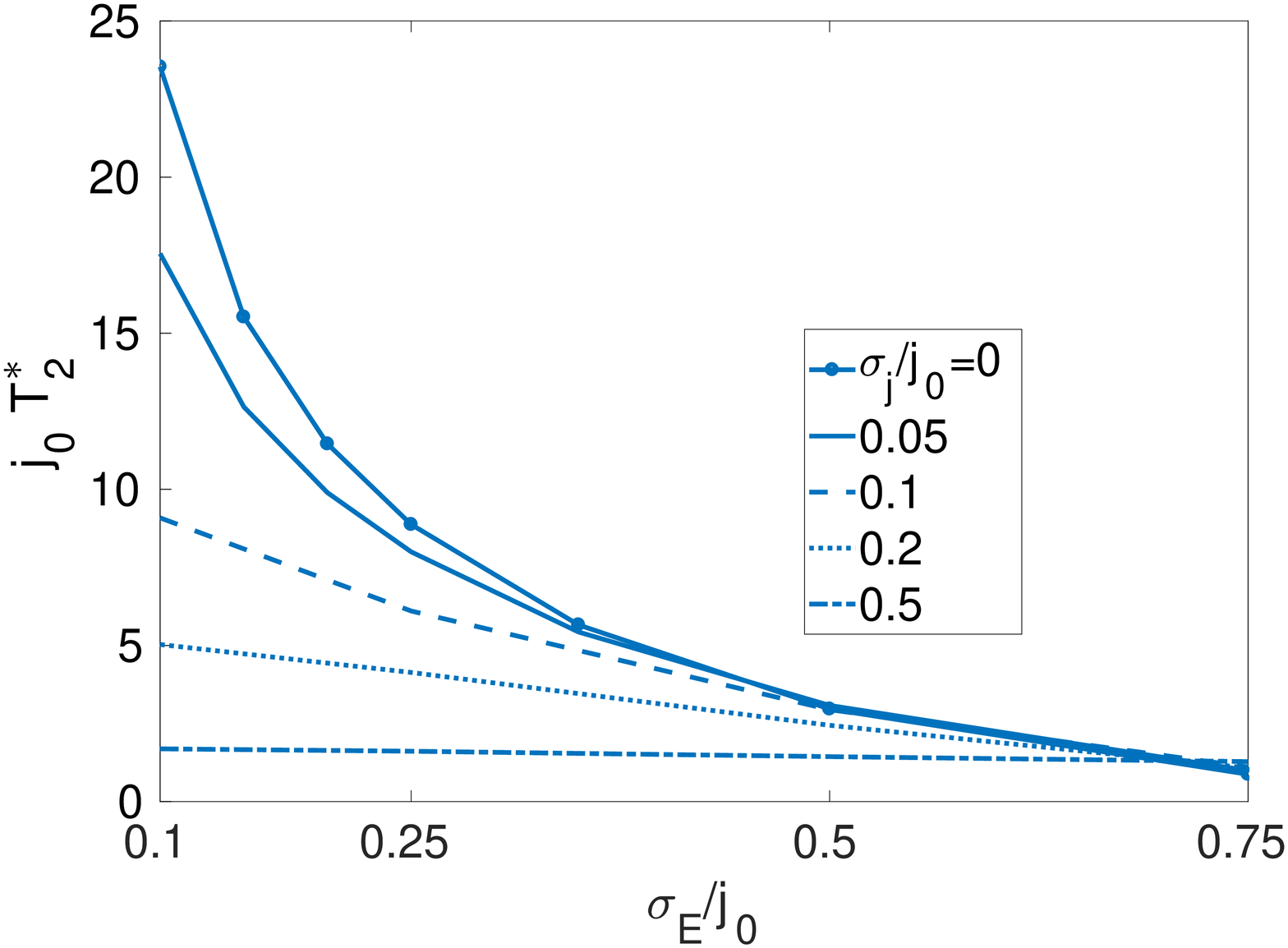}\quad\includegraphics[width=5.5cm]{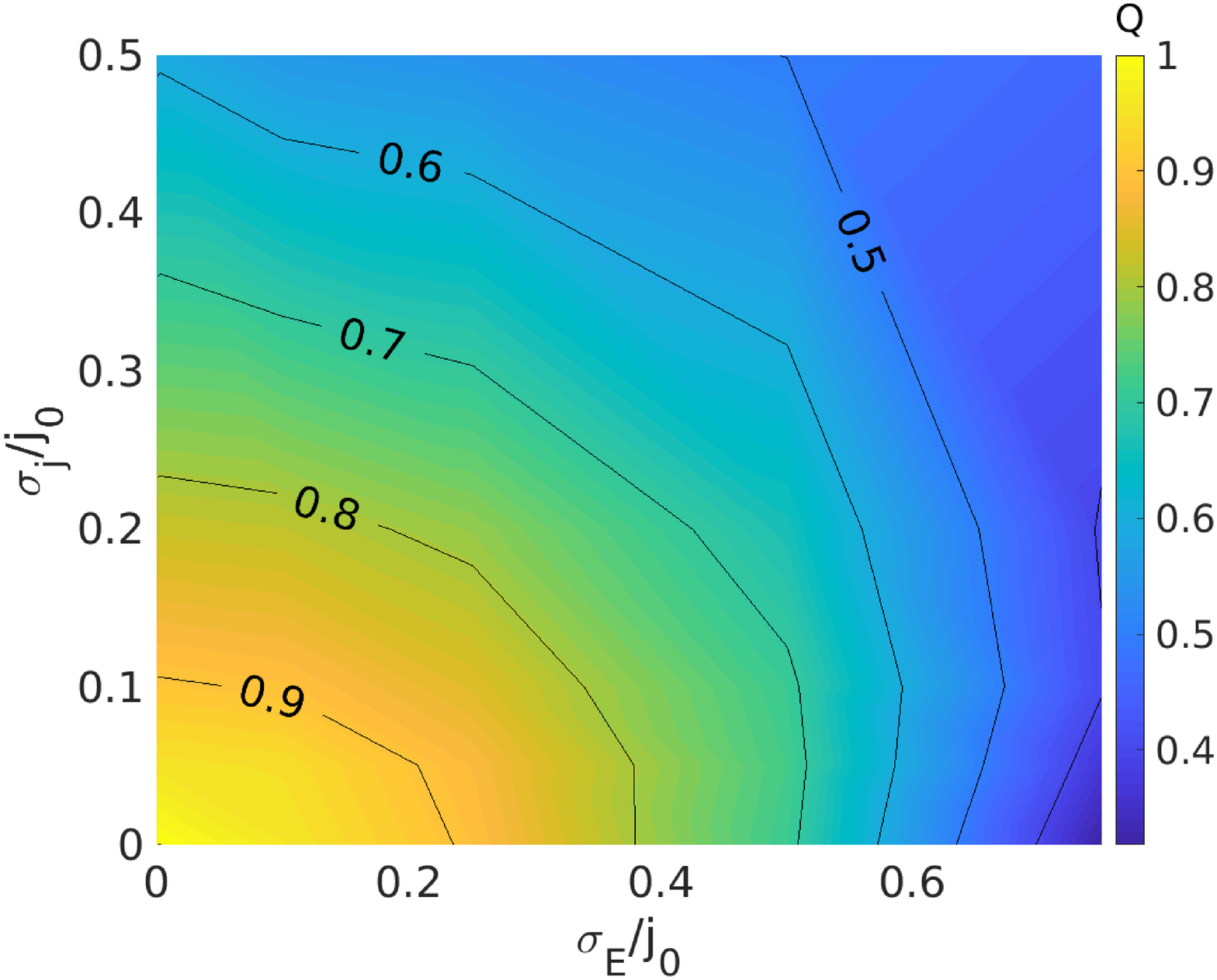}
\caption{Left: Plot of the dimensionless coherence time $j_0 T_2^{\ast}$ as a function of $\sigma_E/j_0$ for different values of $\sigma_j/j_0$. Right: Quality factor as a function of $\sigma_E$ and $\sigma_j$ with black contour lines marking some significant levels of Q.}\label{Q_cond2} 
\end{figure}

\begin{figure}[htbp]
	\begin{center}
		\includegraphics[width=0.55\textwidth]{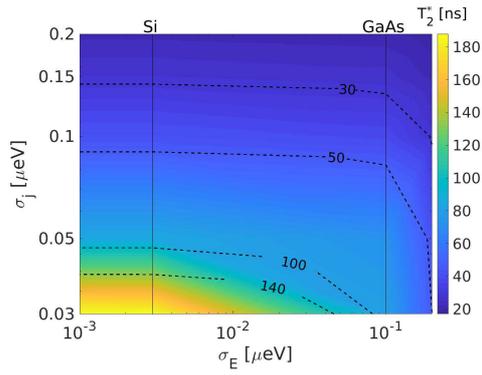}
	\end{center}
	\caption{2D log-log plot of $T_2^{\ast}$ as a function of $\sigma_E$ and $\sigma_j$ when $j_0=1 \mu eV$. The black dashed contour lines mark some levels of $T_{2}^{\ast}$ and vertical solid ones typical minimum magnetic noise variances for Si and GaAs.}\label{3D_cond2} 
\end{figure}

Figs. \ref{Q_cond2} and \ref{3D_cond2} present the effect of charge and magnetic noises on $j_0T_2^{\ast}$ (left) and on $Q$ factor (right). As it can be seen, charge and magnetic noises affect roughly in the same way the coherence time. In addition, it is worth noting that in Figs. \ref{Q_cond2} and \ref{3D_cond2} a more detrimental effect of the noise on the coherence time is observed with respect results reported in Figs. \ref{Q_cond1} and \ref{3D_cond1}. 

While both types of noise suppress $T_{2}^{\ast}$, charge noise is more effective at producing decoherence in the system than magnetic noise. The results presented cover wide parameter regimes for both magnetic and charge noise strengths because the two noise sources have not comparable magnitudes in semiconductors. Our results imply that a given amount of charge noise overall has a more detrimental effect than an equal amount of magnetic noise. 

However, magnetic noise is usually much stronger than charge noise in GaAs, and so this is a problem in this material. Moreover such noise can not be reduced by isotopic purification in GaAs due to the fact that both Ga and As have no stable isotopes with zero nuclear spins. Isotopic purification, on the converse, is exploitable in Si since it can reduce the presence of magnetic isotopes of Si. Another strategy to reduce the effects of magnetic noise that works for all the considered materials is simply to increase the average exchange coupling $j_0$. In fact, the disorder averaged probability depends only on $\sigma_E$/$j_0$ and $\sigma_j$/$j_0$, where $\sigma_E$ remains constant when $j_{0}$ changes and $\sigma_j$ is linear in $j_0$ \cite{Barnes-2016}. 

\subsection{Comparing coherence times in semiconducting DEOQ}
\begin{figure}[htbp]
	\begin{center}
\includegraphics[width=0.6\textwidth]{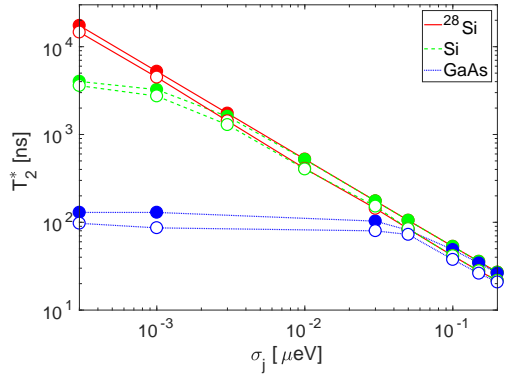}
	\end{center}
	\caption{$T_2^{\ast}$ as a function of $\sigma_j$ when $\sigma_E=0$ $eV$ ($^{28}$Si), 3 n$eV$ (Si) and 100 n$eV$ (GaAs) for the initial conditions $|\psi(0)\rangle=|0\rangle$ (filled marks) and $|\psi(0)\rangle=\frac{1}{\sqrt{2}}(|0\rangle+|1\rangle)$ (open marks) when $j_0=1 \mu eV$.}\label{comparison} 
\end{figure}

Fig. \ref{comparison} collects the most significant results obtained in the previous sections for two initial conditions of interest, the pure logical state $|0\rangle$ (filled marks) and the linear superposition of the two states of the logical basis $\frac{1}{\sqrt{2}}(|0\rangle+|1\rangle)$ (open marks). Addressing three specific values for $\sigma_E$ we report the dependence of $T_2^{\ast}$ in function of $\sigma_j$  when $j_{0}$=1 $\mu$eV. The values addressed for the minimum magnetic noise due to nuclear spins in the host materials are $\sigma_E=0$ eV, 3 neV and 100 neV, that correspond  to the nuclear free isotope $^{28}$Si, the Si and the GaAs, respectively. 
For each host material the simulated coherence time of DEOQ decreases when $\sigma_E$ increases, compressing all curves to similar values when $\sigma_j$ is greater than 0.1 $\mu$eV. When $\sigma_j$ is lower than 0.05 $\mu$eV, the coherence time of DEOQ in GaAs is evidently lower than those of Si and $^{28}$Si whereas $^{28}$Si shows a $T_2^{\ast}$ - in the $\mu$s range - higher than natural Si  only for a smaller $\sigma_j$= 0.003 $\mu$eV. As a result, an implementation of a DEOQ in $^{28}$Si should be preferred to Si only if sufficiently small values of charge noise can be reached. All those considerations holds for both the initial conditions under study.   

If the same level of charge noise is assumed, Si, as well as $^{28}$Si, outperforms GaAs in terms of noise resilience but there are still challenges associated with Si as a material platform. In fact, Si has charge carriers with an higher effective mass than GaAs which require the fabrication of smaller dots to confine single electrons and, moreover, the valley degeneracy of energy levels which impedes the addressing of two nondegenerate levels needed to define the qubit base states.

\section{Conclusions}
The DEOQ dynamical evolution in the "always on" configuration is derived. The purpose of investigating and quantifying how the dynamics is affected by external environmental noise is pursued in view of interesting and fruitful applications that the DEOQ will have in quantum computation. The disorder-average probabilities are calculated starting from two different initial conditions of interest adopting a closed and compact analytical form for the evolution operator that we have derived. The oscillatory dynamical behavior in presence of both magnetic and charge noises is shown and it is demonstrated that the oscillation decay strongly depends on the entity of the disturbance on the qubit. The number of coherent oscillations shown by the return probability before it decays is quantified adopting an envelope-fitting procedure on the disorder-average probabilities through which an estimation of the intrinsic coherence times is obtained. Our results demonstrate that DEOQ is more sensitive to charge noise than magnetic noise and that DEOQ implemented in Si outperforms GaAs qubit in terms of noise resilience. Moreover, only if charge noise is sufficiently small an implementation of DEOQ in $^{28}$Si provides an effective increase of coherence time with respect to Si.  

\section*{Acknowledgments}
This work has been funded from the European Union's Horizon 2020 research and innovation programme under grant agreement No 688539.

%\section*{References}

\bibliographystyle{spphys} 
\bibliography{Ref}

\begin{thebibliography}{10}
\providecommand{\url}[1]{{#1}}
\providecommand{\urlprefix}{URL }
\expandafter\ifx\csname urlstyle\endcsname\relax
  \providecommand{\doi}[1]{DOI \discretionary{}{}{}#1}\else
  \providecommand{\doi}{DOI \discretionary{}{}{}\begingroup
  \urlstyle{rm}\Url}\fi

\bibitem{Shulman-2012}
M.~Shulman, O.~Dial, S.~Harvey, H.~Bluhm, V.~Umansky, A.~Yacoby, Science
  \textbf{336}, 202 (2012)

\bibitem{Veldhorst-2014}
M.~Veldhorst, J.C.C. Hwang, C.H. Yang, A.W. Leenstra, B.~de~Ronde, J.P.
  Dehollain, J.T. Muhonen, F.E. Hudson, K.M. Itoh, A.~Morello, A.S. Dzurak,
  Nature Nanotechnology \textbf{9}, 981 (2014)

\bibitem{Pla-2012}
J.~Pla, K.~Tan, J.~Dehollain, W.~Lim, J.~Morton, D.~Jamieson, A.~Dzurak,
  A.~Morello, Nature \textbf{489}, 541 (2012)

\bibitem{Maune-2012}
B.M. Maune, M.G. Borselli, B.~Huang, T.D. Ladd, P.W. Deelman, K.S. Holabird,
  A.A. Kiselev, I.~Alvarado-Rodriguez, R.S. Ross, A.E. Schmitz, M.~Sokolich,
  C.A. Watson, M.F. Gyure, A.T. Hunter, Nature \textbf{481}, 344 (2012)

\bibitem{Bluhm-2011}
H.~Bluhm, S.~Foletti, I.~Neder, M.~Rudner, D.~Mahalu, V.~Umansky, A.~Yacoby,
  Nature Physics \textbf{7}, 109 (2011)

\bibitem{Tyryshkin-2012}
A.~Tyryshkin, S.~Tojo, J.~Morton, H.~Riemann, N.~Abrosimov, P.~Becker, H.J.
  Pohl, T.~Schenkel, M.~Thewalt, K.~Itoh, S.~Lyon, Nature Material \textbf{11},
  143 (2012)

\bibitem{RuiLi-2012}
R.~Li, X.~Hu, J.~You, Physical Review B \textbf{86}, 205306 (2012)

\bibitem{Coish-2005}
W.~Coish, D.~Loss, Physical Review B \textbf{72}, 125337 (2005)

\bibitem{Shen-2000}
S.Q. Shen, Z.~Wang, Physical Review B \textbf{61}, 9532 (2000)

\bibitem{Morton-2011}
J.J.L. Morton, D.R. McCamey, M.A. Eriksson, S.A. Lyon, Nature \textbf{479}, 345
  (2011)

\bibitem{Kawakami-2014}
E.~Kawakami, P.~Scarlino, D.R. Ward, F.R. Braakman, D.E. Savage, M.G. Lagally,
  M.~Friesen, S.N. Coppersmith, M.A. Eriksson, L.M.K. Vandersypen, Nature
  Nanotechnology \textbf{9}, 666 (2014)

\bibitem{Klymenko-2015}
M.V. Klymenko, S.~Rogge, F.~Remacle, Physical Review B \textbf{92}, 195302
  (2015)

\bibitem{Gamble-2015}
J.K. Gamble, N.T. Jacobson, E.~Nielsen, A.D. Baczewski, J.E. Moussa,
  I.~Montano, R.P. Muller, Physical Review B \textbf{91}, 235318 (2015)

\bibitem{Saraiva-2015}
A.L. Saraiva, A.~Baena, M.J. Calder\'{o}n, B.~Koiller, J. Phys.: Condens.
  Matter \textbf{27}, 154208 (2015)

\bibitem{Loss-1998}
D.~Loss, D.P. DiVincenzo, Physical Review A \textbf{57}, 120 (1998)

\bibitem{DiVincenzo-2000}
D.P. DiVincenzo, D.~Bacon, J.~Kempe, G.~Burkard, , K.B. Whaley, Nature (London)
  \textbf{408}, 339 (2000)

\bibitem{Taylor-2005}
J.~Taylor, H.A. Engel, W.~D\"ur, A.~Yacoby, C.~Marcus, P.~Zoller, M.~Lukin,
  Nature Physics \textbf{1}, 177 (2005)

\bibitem{Laird-2010}
E.~Laird, J.~Taylor, D.~DiVincenzo, C.~Marcus, M.~Hanson, A.~Gossard, Physical
  Review B \textbf{82}, 075403 (2010)

\bibitem{Levy-2002}
J.~Levy, Physical Review Letters \textbf{89}, 147902 (2002)

\bibitem{Petta-2005}
J.R. Petta, A.~C., J.M. Taylor, E.A. Laird, A.~Yacoby, M.D. Lukin, C.M. Marcus,
  M.P. Hanson, A.C. Gossard, Science \textbf{309}, 2180 (2005)

\bibitem{Kikkawa-1998}
J.~Kikkawa, D.~Awschalom, Physical Review Letters \textbf{80}, 4313 (1998)

\bibitem{Amasha-2008}
S.~Amasha, K.~MacLean, I.P. Radu, D.~Zumb\"uhl, M.~Kastner, M.~Hanson,
  A.~Gossard, Physical Review Letters \textbf{100}, 046803 (2008)

\bibitem{Koppens-2008}
F.~Koppens, K.~Nowack, L.~Vandersypen, Physical Review Letters \textbf{100},
  236802 (2008)

\bibitem{Barthel-2010}
C.~Barthel, J.~Medford, C.~Marcus, M.~Hanson, A.~Gossard, Physical Review
  Letters \textbf{105}, 266808 (2010)

\bibitem{Tyryshkin-2003}
A.~Tyryshkin, S.~Lyon, A.~Astashkin, A.~Raitsimring, Physical Review B
  \textbf{38}, 193207 (2003)

\bibitem{Morello-2010}
A.~Morello, J.J. Pla, F.A. Zwanenburg, K.W. Chan, K.Y. Tan, H.~Huebl,
  M.~M\"{o}tt\"{o}nen, C.D. Nugroho, C.~Yang, J.A. van Donkelaar, A.D.C. Alves,
  D.N. Jamieson, C.C. Escott, L.C.L. Hollenberg, R.G. Clark, A.S. Dzurak,
  Nature \textbf{467}, 687 (2010)

\bibitem{Simmons-2011}
C.~Simmons, J.~Prance, B.V. Bael, T.~Koh, Z.~Shi, D.~Savage, M.~Lagally,
  R.~Joynt, M.~Friesen, S.~Coppersmith, M.~Eriksson, Physical Review Letters
  \textbf{106}, 156804 (2011)

\bibitem{Xiao-2010}
M.~Xiao, M.~House, H.~Jiang, Physical Review Letters \textbf{104}, 096801
  (2010)

\bibitem{vandenBerg-2013}
J.~van~den Berg, S.~Nadj-Perge, V.~Pribiag, S.~Plissard, E.~Bakkers, S.~Frolov,
  L.~Kouwenhoven, Physical Review Letters \textbf{110}, 066806 (2013)

\bibitem{Shi-2012}
Z.~Shi, C.B. Simmons, J.R. Prance, J.K. Gamble, T.S. Koh, Y.P. Shim, X.~Hu,
  D.E. Savage, M.G. Lagally, M.A. Eriksson, M.~Friesen, S.N. Coppersmith,
  Physical Review Letters \textbf{108}, 140503 (2012)

\bibitem{Koh-2012}
T.S. Koh, J.K. Gamble, M.~Friesen, M.A. Eriksson, S.N. Coppersmith, Physical
  Review Letters \textbf{109}, 250503 (2012)

\bibitem{Kim-2014}
D.~Kim, Z.~Shi, C.B. Simmons, D.R. Ward, J.R. Prance, T.S. Koh, J.K. Gamble,
  D.E. Savage, M.G. Lagally, M.~Friesen, S.N. Coppersmith, M.A. Eriksson,
  Nature \textbf{511}, 70 (2014)

\bibitem{Kim-2015}
D.~Kim, D.R. Ward, C.B. Simmons, D.E. Savage, M.G. Lagally, M.~Friesen, S.N.
  Coppersmith, M.A. Eriksson, Npj Quantum Information \textbf{1}, 15004 (2015)

\bibitem{Thorgrimsson-2017}
B.~Thorgrimsson, D.~Kim, Y.C. Yang, L.W. Smith, C.B. Simmons, D.R. Ward, R.H.
  Foote, J.~Corrigan, D.E. Savage, M.G. Lagally, M.~Friesen, S.N. Coppersmith,
  M.A. Eriksson, Npj Quantum Information \textbf{3}, 32 (2017).
\newblock \doi{10.1038/s41534-017-0034-2}

\bibitem{Prati-npj}
D.~Rotta, F.~Sebastiano, E.~Charbon, E.~Prati, npj Quantum Information
  \textbf{3}, 26 (2017)

\bibitem{Ferraro-2017}
E.~Ferraro, M.~Fanciulli, M.~{De Michielis}, Quantum Information Processing
  \textbf{16}, 277 (2017).
\newblock \doi{10.1007/s11128-017-1729-1}

\bibitem{DeMichielis-2015}
M.~De~Michielis, E.~Ferraro, M.~Fanciulli, E.~Prati, Journal of Physics A:
  Mathematical and Theoretical \textbf{48}, 065304 (2015)

\bibitem{Ferraro-2014}
E.~Ferraro, M.~De~Michielis, G.~Mazzeo, M.~Fanciulli, E.~Prati, Quantum
  Information Processing \textbf{13}, 1155 (2014)

\bibitem{Jefferson-1996}
J.~Jefferson, W.~H\"ausler, Physical Review B \textbf{54}, 4936 (1996)

\bibitem{DasSarma-2016}
S.~Das~Sarma, R.E. Throckmorton, Y.L. Wu, Physical Review B \textbf{94}, 045435
  (2016)

\bibitem{DasSarma-2017}
R.E. Throckmorton, E.~Barnes, S.~Das~Sarma, Physical Review B \textbf{95},
  085405 (2017)

\bibitem{Wu-2017}
Y.L. Wu, S.~{Das Sarma}, Physical Review B \textbf{96}, 165301 (2017)

\bibitem{Mehl-2015-2}
S.~Mehl, Physical Review B \textbf{91}, 035430 (2015)

\bibitem{Barnes-2016}
E.~Barnes, M.S. Rudner, F.~Martins, F.K. Malinowski, C.M. Marcus, F.~Kuemmeth,
  Physical Review B \textbf{93}, 121407(R) (2016)

\end{thebibliography}

\end{document}